\documentclass[fontsize=12pt,headsepline]{scrartcl}
\setkomafont{date}{\small}
\setkomafont{author}{\normalsize}
\usepackage[mono=false]{libertine}

\usepackage[top=1.25in, bottom=1.25in,inner=3.3cm,outer=2.5cm]{geometry}
\usepackage[latin1]{inputenc}
\usepackage[all]{xy}
\usepackage{amsmath}
\usepackage{amsthm}
\usepackage{amssymb}
\usepackage{bbm}
\usepackage{bm}
\usepackage{caption}
\usepackage{graphicx}
\usepackage{typearea}
\usepackage{empheq}
\usepackage{pstricks,framed}
\usepackage{hyperref}

\hypersetup{colorlinks, breaklinks,citecolor=blue}

\usepackage{newtxmath}
\usepackage{mathrsfs}
\usepackage{simpler-wick}
\usepackage{cancel}

\newlength{\defbaselineskip}
\newcommand{\setlinespacing}[1]%
           {\setlength{\baselineskip}{#1 \defbaselineskip}}

\mathchardef\ordinarycolon\mathcode`\:
\mathcode`\:=\string"8000
\begingroup \catcode`\:=\active
  \gdef:{\mathrel{\mathop\ordinarycolon}}
\endgroup

\newcommand{\one}{{\vvmathbb 1}}

\newcommand{\Tr}{\operatorname{Tr}}

\theoremstyle{plain}
\newtheorem{thm}{Ejercicio}

\theoremstyle{definition}
\newtheorem{definition}[thm]{Definition}
\newtheorem{example}[thm]{Example}

\theoremstyle{remark}
\newtheorem{remark}[thm]{Remark}

\begin{document}
\setlinespacing{1.1}
\title{\Large Entanglement Entropy in Quantum Mechanics: An Algebraic Approach}
\author{A.F. Reyes-Lega\medskip\\
Departamento de F\'{i}sica, Universidad de los Andes,\\
  A.A. 4976-12340, Bogot\'a, Colombia\\
anreyes@uniandes.edu.co
}
\date{}
\maketitle

\begin{abstract}
An algebraic approach to the study of entanglement entropy of quantum systems is reviewed. Starting with a state on a $C^*$-algebra, one can construct a density operator describing the state in the GNS representation state. Applications of this approach to the study of entanglement measures for systems of identical particles are outlined. The ambiguities in the definition of entropy within this approach are then related to the action of unitaries in the commutant of the representation and their relation to modular theory explained.
\end{abstract}

\section{Introduction}
Entanglement entropy has increasingly become an important concept in physics. From black hole physics~\cite{Bombelli1986,Solodukhin2011} to conformal field theory  and statistical mechanics~\cite{Vidal2003}, to quantum information theory~\cite{Nielsen2010}, its prominent role cannot be overemphasized. In the particular case of bipartite quantum systems described by a tensor product structure of the underlying Hilbert space, the information-theoretic content of entanglement entropy is well known~\cite{Janzing2009}. In the case of entanglement measures for systems of identical particles there has been an intense debate during the last  years~\cite{Benatti2020}. In fact,
many proposals have been put forward which try to provide  sensible definitions of
entanglement measures which take into account the fact that a separation of such a system into subsystems does not necessarily entail a factorization of the Hilbert space into a tensor product.
It is the aim of this chapter to review the basic aspects of an approach to this problem that I have pursued for some time, together with A.P. Balachandran and collaborators~\cite{Balachandran2013,Balachandran2013a,Balachandran2013b,Balachandran2020}.
\section{Observables and States}
Quite generally, it is possible to say that any sensible  physical theory should at least provide a definition of what is to be understood as \emph{observable} and what is supposed to be regarded as a \emph{state} of a given system. Results of measurements, then can be thought of as the result of a ``pairing'' between observables and states.

As an example, let us consider the Hamiltonian description of a classical system of $n<\infty$ degrees of freedom. Assuming a phase space of the form $T^*Q$, with $Q$ the configuration space manifold, we can use local coordinates for position ($q^i$) and momentum ($p_i$). Time evolution is given by the solution $(q(t),p(t))$ of Hamilton's equations for a given Hamiltonian $H(q,p)$, which is a smooth function on phase space, $H\in C^\infty(T^*Q)$. In this context we say that the state of the system at a given time $t_0$ is given by the  point $(q(t_0),p(t_0))\in T^*Q$. We are then led to think of points in $T^*Q$ as possible  states of the system. On the other hand, observable quantities like energy, momentum, {\emph{etc.}}, are given by smooth functions on phase space.
The ``pairing'' we alluded to above is given, in this case, by the evaluation of an observable $f \in C^\infty(T^*Q)$ at the point $(q_0,p_0)$ specified by the state. If the number of degrees of freedom becomes significantly large (as is the case in classical statistical mechanics) a statistical approach becomes necessary. In that case, the state of the system is given by a probability distribution $\rho(q,p)$ and the pairing is now given by the evaluation of the expectation value of $f(q,p)$,
\begin{equation}
\langle f\rangle =\int f(q,p) \rho(q,p) d\mu,
\end{equation}
where $\mu$ is the integration (Liouville) measure. Notice that this is a more general notion of state. We can use it to describe, say, the  canonical ensemble, for which $\rho$ is given (up to a normalization factor) by $\exp(-H(q,p)/k_{\mathrm B}T)$. But we can also recover our initial definition of state, by assigning a Dirac delta distribution to the point $(q_0,p_0)$:
\begin{equation}
\rho(q,p)= \delta (q-q_0, p-p_0).
\end{equation}

In summary, we have the following
\begin{itemize}
\item \emph{Observables} are elements of the algebra $C^\infty(T^*Q)$.
\item \emph{States} are positive, normalized functionals of the form
\begin{align}
\omega_\rho:   C^\infty(T^*Q) &\longrightarrow \; \mathbb R\nonumber\\
                f\qquad &\longmapsto  \omega_\rho(f) = \int f(q,p) \rho(q,p) d\mu.
\end{align}
\end{itemize}
 One of the reasons why the algebraic approach to quantum physics is so useful in order to elucidate the probabilistic structure of quantum theory and its relation to its classical counterpart is that the pairing between observables and states runs, essentially, in parallel to what we have just described in the above example. The difference lies in the types of algebras considered which, as a rule, are non-commutative ones in quantum theory. Positivity and normalization of states (defined as functionals with those properties) allow one to construct probability distributions.
  Following the analogy naturally provided by Gelfand duality, the similarities and (most importantly) the differences between classical and quantum physics are made very clear (the reader may consult~\cite{Reyes-Lega2016} for details).

In quantum physics, $C^*$-algebras and von Neumann algebras are the two kinds of algebras which are most relevant. Since every von Neumann algebra is, in particular, a $C^*$-algebra and since in the finite-dimensional case (the one we are interested in this chapter) the distinction disappears, we will only consider $C^*$-algebras. Let us remark, though, that in (local) quantum field theory, it is essential to take into account  those properties that are particular to von Neumann algebras~\cite{Haag1996}.
\begin{definition} A $C^*$-algebra $(\mathcal A,\|\cdot\|,*)$ is an involutive Banach algebra satisfying the following compatibility condition between the involution ``$*$''  and the norm
``$\|\cdot\|$'':
\begin{equation}
\|aa^*\|=\|a\|^2, \qquad \forall a\in \mathcal A.
\end{equation}
\end{definition}
\begin{definition}\label{def:2} A state $\omega$ on a (unital)  $C^*$-algebra $(\mathcal A,\|\cdot\|,*)$ is a positive, normalized linear functional $\omega: \mathcal A \rightarrow \mathbb C$.
\end{definition}
A fundamental result due to Gelfand, Naimark and Segal (the GNS construction~\cite{Haag1996,Werner2000}) allows one to associate  a representation of any $C^*$-algebra $\mathcal A$ (by bounded operators on a Hilbert space) to each state $\omega:\mathcal A\rightarrow\mathbb C$. The main idea behind the construction is to take advantage of the vector space structure of $\mathcal A$ and to use the product in order to define an action of $\mathcal A$ (as an algebra) on itself (but this time regarded as a vector space). In order to endow this vector space with an inner product, one makes use of the fact that $\omega(a^*b)=\overline{\omega(b^*a)}$ in order to define a sesquilinear form $\langle a,b\rangle:= \omega(a^*b)$. Since there might be elements $a\in \mathcal A\setminus\lbrace 0\rbrace$ for which $\omega(a^*a)=0$, it is necessary to take the quotient by the ideal $\mathcal N_\omega$ generated by all such elements. The Hilbert space for the representation is then defined as the completion of the quotient space:
\begin{equation}
\mathcal H_\omega:=\overline{\mathcal A/\mathcal N_\omega}.
\end{equation}
The (bounded) operator representing $a\in \mathcal A$ is the operator $\pi_\omega(a)$ defined through
\begin{equation}
\pi_\omega (a) |[b]\rangle:=|[ab]\rangle.
\end{equation}
It is through the GNS construction that we can make contact with the standard formulation of quantum mechanics in terms of Hilbert spaces and linear operators acting on them.
\section{Entanglement Entropy}
Let us suppose we are given a composite (bipartite) quantum system described by a Hilbert space of the form  $\mathcal H=\mathcal H_A\otimes \mathcal H_B$. If $|\psi\rangle\in \mathcal H$ is a pure state, then it is well known that $|\psi\rangle$ is an entangled state if and only if the von Neumann entropy of either of its reduced density matrices is greater than zero. The reduced density matrices are obtained by partial trace:
\begin{equation}
\rho_A:= \Tr_{\mathcal H_B}(|\psi\rangle\langle\psi|), \qquad \rho_B:= \Tr_{\mathcal H_A}(|\psi\rangle\langle\psi|).
\end{equation}
Using  the Schmidt decomposition one then shows that $S(\rho_A)=S(\rho_B)$ and that, furthermore, this (von Neumann) entropy is zero if and only if the state $|\psi\rangle$ is separable. The problem with systems of identical particles stems from the fact that the underlying Hilbert space is not a tensor product: many-particle states must be either totally symmetric or totally anti-symmetric tensors. This leads to problems when trying to use partial trace or the usual Schmidt decomposition in this context. Some time ago~\cite{Balachandran2013,Balachandran2013a}   we proposed to use the more general notion of \emph{restriction} of a state to a subalgebra instead of partial trace. If $\omega: \mathcal A\rightarrow\mathbb C$ is a state on $\mathcal A$ and $\mathcal A_0\subset \mathcal A$ is a subalgebra of $\mathcal A$, then the restriction of $\omega$ to $\mathcal A_0$ is a state on $\mathcal A_0$, that can be used to study entanglement properties relative to the subsystem decomposition corresponding to this assignment of a subalgebra. The following example (taken from Refs.~\cite{Balachandran2013,Reyes-Lega2016})  is very useful in order to clarify the connection between restrictions and partial traces.
\begin{example}
Let  $\mathcal H=\mathcal H_A\otimes\mathcal H_B\equiv\mathbb C^2\otimes \mathbb C^2$ and take
$\mathcal{A}=M_2(\mathbb{C})\otimes M_2(\mathbb{C})$ as the observable algebra.
Consider now following state vector,
\begin{equation}
|\psi_\lambda\rangle=\sqrt{\lambda} |+,-\rangle +\sqrt{(1-\lambda)} |-,+\rangle,
\end{equation}
where $0\leq\lambda\leq 1$. We can also regard it as an algebraic state $\omega_\lambda$, in
the sense of Definition \ref{def:2}, if we put $\omega_\lambda (O):= \langle\psi_\lambda|O|\psi_\lambda\rangle$, for all $O\in \mathcal A$.
Let now  $\mathcal A_0\subset \mathcal A$ denote  the subalgebra
generated by elements of the form $\alpha\otimes \one_2$, with $\alpha\in M_2(\mathbb{C})$.
 By the definition, the restriction of $\omega_\lambda$ to $\mathcal A_0$ is given by
\begin{equation}
\omega_{\lambda,0}(\alpha\otimes\one_2)=\langle \psi_\lambda| \alpha\otimes \one_2 |\psi_\lambda\rangle=
\lambda\langle +|\alpha|+\rangle +(1-\lambda)\langle -|\alpha|-\rangle.
\end{equation}
On the other hand, the use of partial trace (with
$\rho_A=\Tr_B
|\psi_\lambda\rangle\langle\psi_\lambda|$) leads to
\begin{equation}
\rho_A=\left(\begin{array}{cc} \lambda & 0\\ 0 & 1-\lambda\end{array}\right).
\end{equation}
It is readily checked that $\omega_{\lambda,0}(\alpha\otimes\one_2) =\Tr_{\mathcal H_A} (\rho_A \alpha)$.
\end{example}
As already remarked, the use of partial trace in the case of identical particles will be problematic, whereas the restriction of a state to a subalgebra is always a well-defined operation, and physically sensible. Being a purely algebraic operation, it remains to see how to compute the von Neumann entropy of the reduced (restricted) state. This can be done using the GNS construction. The idea is quite simple. If we restrict a state $\omega$ on $\mathcal A$ to a subalgebra $\mathcal A_0$, then the entropy of $\omega_0:=\omega|_{\mathcal A_0}$ can be computed as follows. Construct the GNS space $\mathcal H_{\omega_0}$ and find a density matrix $\rho_{\omega_0}$
such that
\begin{equation}
\omega_0(\alpha) = \Tr_{\mathcal H_{\omega_0}}(\rho_{\omega_0} \pi_{\omega_0}(\alpha)).
\end{equation}
This can be done by a careful manipulation of the projectors associated with the decomposition of $\mathcal H_{\omega_0}$ into irreducible representations.
Then we can define
\begin{equation}\label{eq:vN-entropy}
S(\omega_{0}):= -\Tr \left(\rho_{\omega_0} \ln \rho_{\omega_0}\right).
\end{equation}

Explicit examples of this procedure, illustrating its usefulness in the definition of an entanglement measure for systems of identical particles, can be found in Refs.~\cite{Balachandran2013,Balachandran2013b,Reyes-Lega2016}.
\section{An Unexpected Symmetry}
The assignment of an entropy to a state $\omega$ through the GNS construction is based on the possibility of finding a density operator $\rho_\omega$ acting on the representation space $\mathcal H_\omega$ in such a way that
\begin{equation}
 \label{eq:density-matrix}
\omega(a) =\Tr_{\mathcal H_{\omega}}(\rho_{\omega} \pi_{\omega}(a))\quad
\mbox{for all}\quad a \in \mathcal A.
\end{equation}  The relationship between $\omega$ and $\rho_\omega$ is, however, not unique. The choice that was made in~\cite{Balachandran2013}
was based on the  idea of using the projectors $P^{(k)}$ arising from the decomposition of $\mathcal H_\omega$ into irreducible subspaces, $\mathcal H_\omega=\bigoplus_k \mathcal H_\omega^{(k)}$, in order to define
\begin{equation}
\label{eq:omega-rho}
\rho_\omega := \sum_k P^{(k)}|[\one_{\mathcal A}]\rangle\langle [\one_{\mathcal A}] |P^{(k)}.
\end{equation}
This choice might  be regarded as a ``natural'' one, but it is not the only one.
In fact, since the decomposition of $\mathcal H_\omega$ into irreducible subspaces is in general not unique (this can only happen if each irreducible component appears with multiplicity at most one), there are many choices of projectors, which in turn lead to different choices of density operators compatible with (\ref{eq:omega-rho}). The ensuing entropy ambiguities were studied in Refs.~\cite{Balachandran2013e,Balachandran2013d}. In particular, in Ref.~\cite{Balachandran2013d} it was suggested that the ambiguity could be naturally explained in the context of Tomita--Takesaki modular theory. Although the natural context of modular theory is that of von Neumann algebras, an analysis of the entropy ambiguity for finite dimensional algebras using modular theory really sheds light on the problem. Such an analysis, valid for any finite dimensional $C^*$-algebra, has been carried out
 in~\cite{Balachandran2020}. In the following, I will present a simple example~\cite{Balachandran2020,Tabban2022} that illustrates the  general idea.
\begin{example} Let $\mathcal A= M_n(\mathbb C)$ and consider the state $\omega$ defined through
\begin{equation}
\omega(a):= \sum_{i}^n \lambda_i a_{ii},
\end{equation}
where the coefficients $\lambda_i$ are assumed to be strictly positive and $\sum_{i=1}^n \lambda_i=1$. The fact that $\lambda_i>0$ for all $i$ implies that there is no null space ($\mathcal N_\omega = \lbrace 0 \rbrace$), so that vectors in $\mathcal H_\omega$ have only one representative. We may, therefore, write simply $|a\rangle$ to denote elements in $\mathcal H_\omega$. The inner product is given by $\langle a | b \rangle = \omega (a^*b)$.
As $\omega(a)= \langle \one_n|\pi_\omega(a)|\one_n\rangle$, we see that the state vector $|\one_n\rangle$ gives us a \emph{purification} of the state $\omega$. The algebra $\mathcal A$ (system $\mathbf A$) is now faithfully represented as a subalgebra $\pi_\omega(\mathcal A)$ of $\mathcal L (\mathcal H_\omega)$ --an emergent system we will denote as system $\mathbf C$-- and which in turn contains the commutant $\pi_\omega(\mathcal A)'$ of $\pi_\omega(\mathcal A)$, denoted here as system $\mathbf B$. According to modular theory, observables in $\mathbf B$ are obtained from observables in $\mathbf A$ by the action of the modular conjugation $J$~\cite{Tabban2022}. Indeed, if $\alpha \in \pi_\omega (\mathcal A)$, then $J\alpha J\in \pi_\omega (\mathcal A)'$. In our simple example, $J$ can be given described explicitly in terms of its action on a basis of matrix units $e_{ij}$ for $\mathcal A$:
\begin{equation}
J|e_{ij}\rangle = \sqrt{\lambda_j/\lambda_i} |e_{ji}\rangle.
\end{equation}
We refer the reader to~\cite{Balachandran2020, Tabban2022} for details. Let now $|\hat e_{ij}\rangle$ denote the vector $|e_{ij}\rangle$, properly normalized with respect to the GNS inner product. It follows that the projectors entering Eq. (\ref{eq:omega-rho}) can in this case be written as follows:
\begin{equation}
P^{(k)}=\sum_{i=1}^n |\hat e_{ik}\rangle \langle\hat e_{ik}|.
\end{equation}
As remarked above, any other orthogonal decomposition of $\mathcal H_\omega$ providing an equivalent representation of $\mathcal A$ will lead to a density operator satisfying (\ref{eq:density-matrix}). In particular, given any unitary element of $\mathcal A$ ($g\in \mathcal A,\; g^*g=\one_n$), we can induce an equivalent decomposition by defining
$|\hat e_{ik}(g)\rangle:= U(g)|\hat e_{ik}\rangle$, where
\begin{equation}
U(g):= J\pi_\omega (g)J\in \pi_\omega (\mathcal A)'
\end{equation}
is a unitary operator that belongs to the commutant and which, therefore, can be regarded as a \emph{gauge transformation}.
\end{example}
Let us denote with  $\rho_\omega (g)$ the density operator obtained by replacing the projectors $\lbrace P^{(k)}\rbrace_{k=1,\ldots,n}$ in (\ref{eq:omega-rho}) by the following ones,
\begin{equation}
P_g^{(k)}:=J \pi_\omega(g p^{(k)} g)J,\quad k=1,\ldots,n,
\end{equation}
where $g$ is a unitary element of $\mathcal A$ and $p^{(k)}:= e_{kk}$. It can be readily shown that $\rho_\omega (g)$ still satisfies (\ref{eq:density-matrix}) and can be used to describe the original state $\omega$ in system $\mathbf A$. But its entropy will in general differ from that of $\rho_\omega \equiv \rho_\omega (\one_n)$. In fact, we have ({\emph{cf.}} Theorem 2 in~\cite{Balachandran2020}):
\begin{equation}
S(\rho_\omega (g)) \ge S(\rho_\omega).
\end{equation}
The above example can be generalized to the case of arbitrary finite dimensional $C^*$-algebras~\cite{Balachandran2020}. The emergent gauge symmetry appearing through the action of the unitaries $U(g)$ has been related to the action of the gauge group of the fiber bundle describing ``molecular shapes'' in~\cite{Balachandran2013d}. A generalization of this approach to the case of homogeneous spaces of compact Lie groups has been studied by Tabban in~\cite{Tabban2022}.
\begin{remark} Recently, an alternative approach to the study of the entropy ambiguities discussed in this section has been proposed by Facchi and collaborators~\cite{Facchi2021}. It would be interesting to relate the two approaches.
\end{remark}
\section{Working with Bal}
I met Bal for the first time during a conference on the connection between spin and statistics that took place in Trieste back in 2008. During the last decade I have had the pleasure and the honor of collaborating with him. The lessons learnt from his way of doing research are countless, but there is one which has been particularly important for me and my group in Bogot{\'a}, namely the group discussions. Although I did not take part in the legendary ``Room-316 meetings'', I was lucky enough to enjoy many discussions on different occasions that I adopted as part of my way of doing research with my students, some of whom are now my younger collaborators. Seeing how fast they can learn --and become accomplished researchers-- has shown me the value that a generous and lively exchange of ideas (something that Bal has always fostered) can have.


\end{document}